\documentclass[nofootinbib,reprint,groupedaddress,preprintnumbers]{revtex4-2}

\usepackage[bottom]{footmisc}
\usepackage{amsmath,amssymb,marginnote}
\usepackage{multirow} 
\usepackage{booktabs} 
\usepackage{graphicx,subfigure}
\usepackage{dcolumn}
\usepackage{bm}
\usepackage[mathlines]{lineno}
\usepackage[colorlinks=true, allcolors=blue]{hyperref}
\usepackage[usenames]{color}
\usepackage{xcolor}
\usepackage{ulem}
 
\begin{document}

\preprint{LA-UR-25-30145}

\title{Testing common approximations of neutrino fast flavor conversion}

\author{Erick Urquilla}
\email{eurquill@vols.utk.edu}
\affiliation{Department of Physics and Astronomy, University of Tennessee, Knoxville, TN 37996, USA}

\author{Lucas Johns}
\email{ljohns@lanl.gov}
\affiliation{Theoretical Division, Los Alamos National Laboratory, Los Alamos, NM 87545, USA}

\begin{abstract}

    A new chapter is opening in the theory of core-collapse supernovae and neutron star mergers as simulations of these events begin to incorporate fast flavor conversion (FFC) and other forms of neutrino flavor mixing. Using numerical experiments, we show that the approximations of FFC that have been implemented so far are limited by at least two of three factors: (1) approximating continuous evolution as a discrete sequence of instabilities, (2) using spatially homogeneous asymptotic states, and (3) assuming that FFC must be accompanied by instability. The factors we identify in this work will be important considerations as the research area progresses from initial exploratory studies to more quantitatively precise assessments.

\end{abstract}

\maketitle

\section{Introduction}
    
    Neutrino oscillations in core-collapse supernovae and neutron star mergers pose an outstanding challenge at the forefront of particle astrophysics \cite{volpe2024neutrinos, johns2025neutrino, tamborra2025neutrinos}. A growing body of evidence indicates that the astrophysical effects of flavor mixing, particularly of FFC, may be quite significant \cite{sawyer2016neutrino, wu2015effects, wu2017fast, 
    wu2017imprints, abbar2019occurrence, delfan2019linear, morinaga2020fast, xiong2020potential, ko2020neutrino, george2020fast, abbar2021characteristics, nagakura2021where, tamborra2021new, li2021, just2022, fernandez2022, fujimoto2023explosive, xiong2023collisional, ehring2023, ehring2023fast, ehring2024gravitational, akaho2024collisional, liu2024muon, mukhopadhyay2024time, qiu2025neutrino, wang2025, lund2025angle, mori2025, Nagakura_2019}. The effects of slow \cite{kostelecky1993neutrino, duan2006collective, duan2010collective, dasgupta2015temporal, chakraborty2016collective, shalgar2024neutrino, fiorillo2025theoryslow, fiorillo2025theoryslow2, fiorillo2025first} and collisional \cite{johns2021collisional, johns2022collisional, xiong2022evolution, xiong2022collisional, liu2023systematic, liu2023universality, akaho2023collisional, shalgar2023neutrinos, kato2023collisional, froustey2024neutrino, zaizen2025spectral, froustey2025predicting, wang2025effect} flavor conversion are currently less clear but possibly considerable as well.
    
    It is widely hoped that a viable solution to the problem will involve coarse-graining over the small spatial scales associated with flavor mixing \cite{li2021, zaizen2023simple, shalgar2023neutrino, shalgar2024length, xiong2023evaluating, xiong2024robust, abbar2024physics, richers2024asymptotic, nagakura2024bhatnagar, johns2023thermodynamics, johns2024subgrid, johns2025local, johns2025implications, kost2024once, kost2025once, fiorillo2024fast, fiorillo2025collective, liu2024quasi, liu2025resolution, liu2025asymptotic, liu2025dynamical, gdg9-rzns}. Two classes of FFC subgrid models have already been implemented into astrophysical simulations: effective classical transport \cite{li2021, xiong2024robust, just2022, fernandez2022, ehring2023, lund2025angle, mori2025} and the Bhatnagar--Gross--Krook (BGK) subgrid model \cite{nagakura2024bhatnagar, qiu2025neutrino, wang2025, wang2025effect}. In this work we present the results of numerical tests that highlight some crucial limitations of these two methods. 
    
    In principle, to identify the deficiencies of any approximate treatment of flavor mixing, one would need to solve the full quantum kinetic equation (QKE), including incoherent collision terms and the processes that drive the system toward instability. Those results would then provide a natural baseline for comparison. However, performing this test in fully self-consistent global QKE radiation-hydrodynamics simulations, where electron lepton number minus heavy lepton number (ELN-XLN) angular crossings arise naturally from advection and incoherent neutrino interactions with the bulk fluid, is computationally challenging. We therefore replace this ideal baseline with a controlled alternate procedure: we simulate a centimeter-scale domain in which the ELN-XLN angular crossing is artificially induced through neutrino injection. In this restricted sense, injection serve as a controlled alternative for the astrophysical mechanisms that produce ELN-XLN crossings in global simulations, allowing us to examine how a crossing influences the system’s evolution toward an asymptotic state in a computationally tractable setting.
    
    Effective classical transport is based on a two-step procedure. In the first step, the astrophysical simulation is advanced without neutrino oscillations. Flavor instabilities emerge in some regions during this part of the procedure. Then, in the second step, those newly formed unstable regions are set to the corresponding post-instability asymptotic states (or, in many instances, some very rough approximation of those states). The argument for this two-step method is that the time scale for a fast instability to bring the local neutrino flavor distributions to an asymptotic state is allegedly much shorter than the time scale on which astrophysical conditions are changing \cite{li2021}.
    
    However, this method does not self-consistently apply the assumed scale separation \cite{johns2024subgrid}. The reason fast instabilities emerge during the first step of the procedure is that flavor mixing---the means by which neutrinos naturally try to suppress the emergence of instabilities---has been artificially shut off. Effective classical transport might nonetheless be vindicated despite the lack of self-consistency if, in the limit of infinitesimal step size, the method converges on the true continuous evolution. This possibility is not particularly comforting, however, unless convergence already occurs at relatively large step size.
    
    Reference~\cite{fiorillo2024fast} compared the ultimate flavor outcomes when neutrinos are injected inside a periodic box either slowly and continuously or all at once generating ELN--XLN crossings and FFC (see also Ref.~\cite{liu2024quasi}). In some cases the outcomes were found to agree, in other cases not. These comparisons can be interpreted as tests of the two-step aspect of effective classical transport. Sudden-injection asymptotic states are used in the second step of effective classical transport, and so comparing them to continuous-injection asymptotic state gauges the amount of error introduced in each location, at each time step, by using a stepwise procedure. Our first goal is to highlight this implication of such comparison calculations and present additional evidence that the error from approximating continuous evolution as a sequence of instabilities is a concern for effective classical transport (Sec.~\ref{sec:sequential}). The findings of Sec.~\ref{sec:sequential} reinforce the critique of Ref.~\cite{johns2024subgrid} and motivate further investigation into why the sequential-instability treatment works when it does.
    
    In Sec.~\ref{sec:homog} we point out a second and perhaps more serious limitation of some subgrid methods. All asymptotic-state prescriptions devised so far are spatially homogeneous at subgrid scales (e.g., \cite{zaizen2023simple, xiong2023evaluating}). We show that the imposition of subgrid homogeneity engenders even greater discrepancies with the true evolution than the two-step approximation does, as evidenced by calculations with continuous injection of neutrinos but periodic homogenization. The BGK subgrid model avoids the continuous/sequential issue of Sec.~\ref{sec:sequential}, but if it prescribes local relaxation to a spatially homogeneous asymptotic state---as has always been the case so far---then it, like effective classical transport, is subject to the homogenization issue of Sec.~\ref{sec:homog}.
    
    From the standpoint of coarse-grained flavor-wave transport, which has been developed in the contexts of quasilinear theory \cite{fiorillo2024fast, fiorillo2025collective} and miscidynamics \cite{johns2023thermodynamics, johns2024subgrid, johns2025local}, homogeneous BGK models continuously and artificially deplete the spectrum of flavor waves (\textit{i.e.}, the emergent collective degrees of freedom associated with subgrid inhomogeneity). In effective classical transport with homogeneous asymptotic states, the flavor-wave spectrum is in effect periodically replaced by small fluctuations. As an alternative to coarse-grained flavor-wave transport, one could attempt to formulate spatially inhomogeneous asymptotic states for use in effective classical transport or the BGK model. This idea has not yet been developed, nor do we do so here.
    
    In Secs.~\ref{sec:irrev} and \ref{sec:random} we explore some other aspects of subgrid inhomogeneity. Section~\ref{sec:irrev} makes a point that, to our knowledge, has not been made before: FFC does not need to be accompanied by instability. We demonstrate this in the following manner. First, neutrinos are slowly injected in such a way that the system is driven through marginally stable states. During this phase, fast instabilities cause subgrid inhomogeneity to build up. We then switch to slowly removing neutrinos in such a way that the system is driven through (non-marginally) stable states and encounters no further instabilities. FFC occurs due to the preexisting inhomogeneity despite the absence of instabilities.
    
    We set up our calculations in Sec.~\ref{sec:irrev} so that, if flavor mixing were turned off, the removal of neutrinos would undo the injection and return the system to the initial state. In the exact solution, the flavor-mixing system nearly returns to its initial state but not quite. The degree of irreversibility is much more pronounced in calculations where the subgrid inhomogeneities are replaced by small perturbations at the moment injection switches to removal. Effective classical transport and the BGK model, at least in the forms implemented so far (in particular, with local relaxation to post-instability asymptotic states), equate FFC to unstable evolution. This is the third limitation that we highlight.
    
    One of the main messages of the paper is that subgrid flavor inhomogeneities are generally important to keep track of in astrophysical settings. Still, one would hope that some of the subgrid details prove to be inessential. If not, the viability of all coarse-grained methods is questionable. One particularly optimistic idea is to approximate the evolution in terms of $\langle \bm{P} \rangle$ and $\langle |\bm{P}| \rangle$, where $\bm{P}$ is a polarization vector and $\langle \cdot \rangle$ is a spatial coarse-graining operator \cite{liu2024quasi}. This is a quasi-homogeneous analysis in the sense that $\langle |\bm{P}| \rangle$, although a spatially averaged quantity, is sensitive to subgrid information yet still has better predictability properties than the chaotic nature of the subgrid flavor structure \cite{urquilla2024chaos}.
    
    Section~\ref{sec:random} is inspired by the coarse-grained approach of Ref.~\cite{liu2024quasi}, in which only a minimal amount of subgrid information is retained. We present calculations in which we periodically randomize the phases of the transverse polarization vectors $\bm{P}_{\bot}(z)$ (that is, transverse to the flavor axis) at each location and direction. This procedure leaves the quasihomogeneous quantities unaltered, allowing us to assess whether the subgrid phases matter. We observe some amount of success, but also some alarming discrepancies. Subgrid phases cannot generally be disregarded.
    
    Section~\ref{sec:summary} summarizes our findings. The past few years have been a period of rapid advance in understanding the astrophysical consequences of neutrino flavor mixing at extreme densities. The subgrid methods we put to the test in this paper have been instrumental in that progress. By giving concrete examples of the limitations of these methods, we hope our results will spur the development of further improvements.

\section{Model and methods}
    
    \subsection{Quantum kinetic equation}
        
        We solve an inhomogeneous four-beam neutrino field under quantum kinetic evolution using the \textsc{Emu} code~\cite{richers2021particle}. \textsc{Emu} is a particle-in-cell neutrino transport code designed to model flavor evolution in astrophysical settings such as supernovae or binary neutron star mergers. In \textsc{Emu}, the state of a neutrino field is described by a seven-dimensional $2\times2$ matrix distribution function $f_{ab}(t, \mathbf{x}, \mathbf{p})$. In this work we limit our scope to two neutrino flavors: electron and heavy. The heavy flavor represents a combination of the muon and tau neutrinos. The diagonal components represent occupation numbers, while the off-diagonal components encode flavor coherence. The distribution function matrix and the number density matrix are related by
        \begin{align*}
        n_{ab}(t,\,\mathbf{x}) & = \frac{1}{(hc)^3} \int \, d^3\mathbf{p} \, f_{ab}(t,\,\mathbf{x},\,\mathbf{p}).
        \end{align*}
        
        The time evolution of $f_{ab}$ is governed by the quantum kinetic equation (QKE):  
        \begin{equation}
            (\partial_t+\mathbf{v}\cdot \nabla_{x}) f_{ab} = - \frac{i}{\hbar}[H, f]_{ab} + C_{ab}.
            \label{eq:QKE}
        \end{equation}
        In the massless fast limit, the dominant contribution to the Hamiltonian $H_{ab}$ arises from the neutrino–neutrino forward-scattering potential,  
        \begin{equation}
        \begin{aligned}
        H^{\nu-\nu}_{ab}(t, \mathbf{x}, \mathbf{p}) 
        &= \frac{\sqrt{2}\, G_F}{(h c)^3}
           \int d^3\mathbf{q}\,
           (1 - \hat{\mathbf{p}} \cdot \hat{\mathbf{q}} ) \\[4pt]
        &\quad\times 
           \left[f_{ab}(t, \mathbf{x}, \mathbf{q})
           - \bar{f}^{*}_{ab}(t, \mathbf{x}, \mathbf{q})\right],
        \end{aligned}
        \label{nunuham}
        \end{equation}
        where bar quantities represent antiparticles and $^*$ the complex conjugate. 
        
        In this work, we neglect the physical incoherent collision terms most commonly included in global astrophysical simulations, such as neutrino–nucleon absorption and emission, pair annihilation, and neutrino scattering on baryons and leptons. Instead, we introduce an artificial collision term designed solely to mimic, in a controlled way, the generation of an ELN-XLN crossing. This term injects and removes neutrinos, but it does not model collisional decoherence of the neutrino field and it does not produce collisional flavor instabilities. 
    
        The explicit form of the collision term is specified in terms of $\dot{n}$ in the subsequent sections where it is introduced and used. The relation between $\dot{n}$ and $C_{ab}$ is
        \begin{equation}
        \dot{n}_{ab}(t,\mathbf{x})
        = \frac{1}{(h c)^3}\int \mathrm{d}^3\mathbf{p}\; C_{ab}(t,\mathbf{x},\mathbf{p}) .
        \label{eq:ndot_c}
        \end{equation}
        We use an angular-dependent collision term expressed in terms of $\dot{n}_{ab}^{\mathrm{axis}}$, where $\mathrm{axis}\in\{+x,-x,+z,-z\}$ for our four-beam setup. Each $\dot{n}_{ab}^{\mathrm{axis}}$ represents the contribution from the phase-space associated with that beam direction. That is, $\dot{n}_{ab}^{\mathrm{axis}}$ is given by Eq.~\eqref{eq:ndot_c} with integration limits over a solid angle of $\Omega=\pi$ pointing on the corresponding axis.

    \subsection{Bhatnagar--Gross--Krook approximation\label{sec:bgk}}
    
        The BGK subgrid model \cite{nagakura2024bhatnagar, Akaho_2025} replaces the commutator in the QKE (Eq.~\ref{eq:QKE}) with a relaxation term that drives the system toward an asymptotic state $f^a$ on a timescale $\tau$
        \begin{eqnarray}
        - \frac{i}{\hbar}[H, f] \;\;\longrightarrow\;\; -\frac{1}{\tau}\,(f-f^a).
        \end{eqnarray}
        
        To evaluate the relaxation factor $\tau$ and the asymptotic flavor state $f^a$, we adopt the approach of Ref.~\cite{liu2025asymptotic}. We first introduce the ELN-XLN angular distribution  
        \begin{equation}
            G \equiv \sqrt{2}\, G_F \Big[ (f_e - \bar{f}_e) - (f_{x} - \bar{f}_{x}) \Big],
            \label{eq:G}
        \end{equation}
        and divide the solid-angle integration into two complementary regions, $G<0$ and $G>0$:  
        \begin{align}
            A &\equiv \int_{G < 0} \frac{d\Omega}{4\pi}\, G, \label{eq:A}\\
            B &\equiv \int_{G > 0} \frac{d\Omega}{4\pi}\, G. \label{eq:B}
        \end{align}
        
        From these quantities, the relaxation timescale is defined as  
        \begin{equation}
            \tau = \frac{2\pi}{\sqrt{A B}}.
            \label{eq:tau}
        \end{equation}
        
        The asymptotic flavor distributions are then written in terms of a survival probability $P$:  
        \begin{align}
            f_{ee}^a &= P f_{ee} + (1-P)\, f_{xx},\\
            f_{xx}^a &= (1-P)\, f_{ee} + P f_{xx}.
        \end{align}
        The same equations hold for antineutrinos, with all quantities replaced by their barred counterparts. However, in this work, we do not simulate antineutrinos. $P$ takes the form  
        \begin{align}
            P &=
            \begin{cases}
                \tfrac{1}{2}\left(1 - \tfrac{A}{2B}\right), & G < 0, \\
                \tfrac{1}{2}, & G > 0,
            \end{cases}
            \label{eq:Pij1}
        \end{align}
        for $B \geq A$, and  
        \begin{align}
            P &=
            \begin{cases}
                \tfrac{1}{2}, & G > 0, \\
                \tfrac{1}{2}\left(1 - \tfrac{B}{2A}\right), & G < 0,
            \end{cases}
            \label{eq:Pij2}
        \end{align}
        otherwise.  
        
        This prescription ensures conservation of the total lepton number along each propagation direction, as well as conservation of the total number density.
    
    \subsection{Numerical Setup \label{sec:numsetup}}
    
        As a baseline, we analyze two neutrino configurations that we call the sudden cases. We employ a four-beam setup to give the system greater flexibility in how flavor is distributed over angle, unlike the two-beam model, which is tightly constrained by conserved quantities. In both sudden cases, beams propagating along $\pm x$ carry an electron neutrino number density of $2.45\times 10^{32}\,\mathrm{cm}^{-3}$, and beams along the $+z$ direction carry $4.89\times 10^{32}\,\mathrm{cm}^{-3}$. To generate a fast flavor instability, we create an ELN-XLN crossing by injecting heavy neutrinos propagating in the $-z$ direction with a density of either $2.45\times 10^{32}\,\mathrm{cm}^{-3}$ or $1.96\times 10^{33}\,\mathrm{cm}^{-3}$ (eight times larger than in the first case). The blue curves in Fig.~\ref{fig:seqins} display the flavor evolution paths for the former (upper panel) and the later (lower panel) sudden cases. We exclude antineutrinos from all simulations.
        
        Other simulations modify the sudden cases by changing how heavy neutrinos are injected into or removed from the beam propagating in the $-z$ direction. It is important to note that the injection and removal of heavy neutrinos do not represent a collision term associated, for example, with neutrino-nucleon absorption and emission, pair annihilation, or neutrino scattering on fluid baryons and leptons. Nevertheless, the artificial injection serves the sole purpose of reproducing a smooth mechanism for the generation of ELN-XLN angular crossings. This neutrino injection does not play the role of collisions, since it does not decohere the flavor fields or produce collisional instabilities, but it effectively reproduces the smooth generation of ELN-XLN crossings present in astrophysical scenarios.
        
        We simulate a box of $1\times 1\times 50\,\mathrm{cm}$ resolved by a grid of $1\times 1\times 5000$ cells. The four beams are initialized from the center of each cell. The Nyquist limit wavelength of $0.02\,\mathrm{cm}$ enables us to confidently resolve flavor waves larger than $0.1\,\mathrm{cm}$ and smaller than $50\,\mathrm{cm}$. We impose periodic boundary conditions to preserve homogeneity in the $x$ and $y$ directions and to allow flavor inhomogeneities only along $z$. We neglect vacuum and matter effects. We seed the off-diagonal components of the distribution function with random perturbations whose amplitudes are four orders of magnitude smaller than those of the diagonal components.  
        
        Subsequent figures present the number density evolution in terms of the $z$ component of the SU(2) flavor polarization vector,
        \begin{align}
            P_z &= \frac{n_{ee} - n_{xx}}{n^{+z}_{\mathrm{init}}}.
        \end{align}
        We normalize $P_z$ by 
        \begin{eqnarray}
            n^{+z}_{\mathrm{init}} = 4.89\times 10^{32}\,\mathrm{cm}^{-3},    
        \end{eqnarray}
        the initial density of neutrinos propagating in the $+z$ direction. 
        
        All figures report time in units of the neutrino self-interaction timescale,
        \begin{equation}
        \mu^{-1} = 4.25\times 10^{-12}\,\mathrm{s} \, .
        \end{equation}
        Here $\mu=\sqrt{2}\,G_F\,n$, were we adopt the total initial number density from the sudden simulation shown in the upper panel of Fig.~\ref{fig:seqins}.

\section{Sequential instability\label{sec:sequential}}

    In this section, we examine how the instability-driving scheme, whether sudden, discrete, or continuous, affects the final asymptotic state, with the goal of testing two-step effective classical-transport subgrid models. In these models, one first evolves the classical neutrino radiation field without flavor conversion, thereby allowing instabilities to develop, and introduces flavor conversion only afterward by resetting unstable regions to their post-instability asymptotic states. The discrete-injection setup is designed to mimic this two-step approximation: the discrete injection that produces ELN--XLN crossings represents the classical evolution driving the system into instability, after which FFC is activated through full QKE evolution and the system is allowed to relax naturally. We use full QKE evolution, rather than imposing a prescribed post-instability state, so that we can isolate the error intrinsic to the two-step procedure itself from the additional error associated with choosing a particular asymptotic state. By comparing the asymptotic states obtained with discrete and continuous injection, we quantify the error introduced by the two-step approximation. In realistic astrophysical environments, collisional processes such as emission, absorption, and scattering, together with the flavor-mixing response they induce, act simultaneously within the QKE. By contrast, the two-step approximation replaces this continuous evolution with a stepwise sequence in which the system is first driven without flavor mixing and only afterward allowed to undergo flavor conversion without further driving. The sudden and discrete injection schemes reproduce this stepwise character because the injection is instantaneous, leaving no time for flavor mixing to respond during the driving itself, and is followed by intervals of flavor evolution with no additional injection.
    
\begin{figure}
    \includegraphics[width=1.0\linewidth]{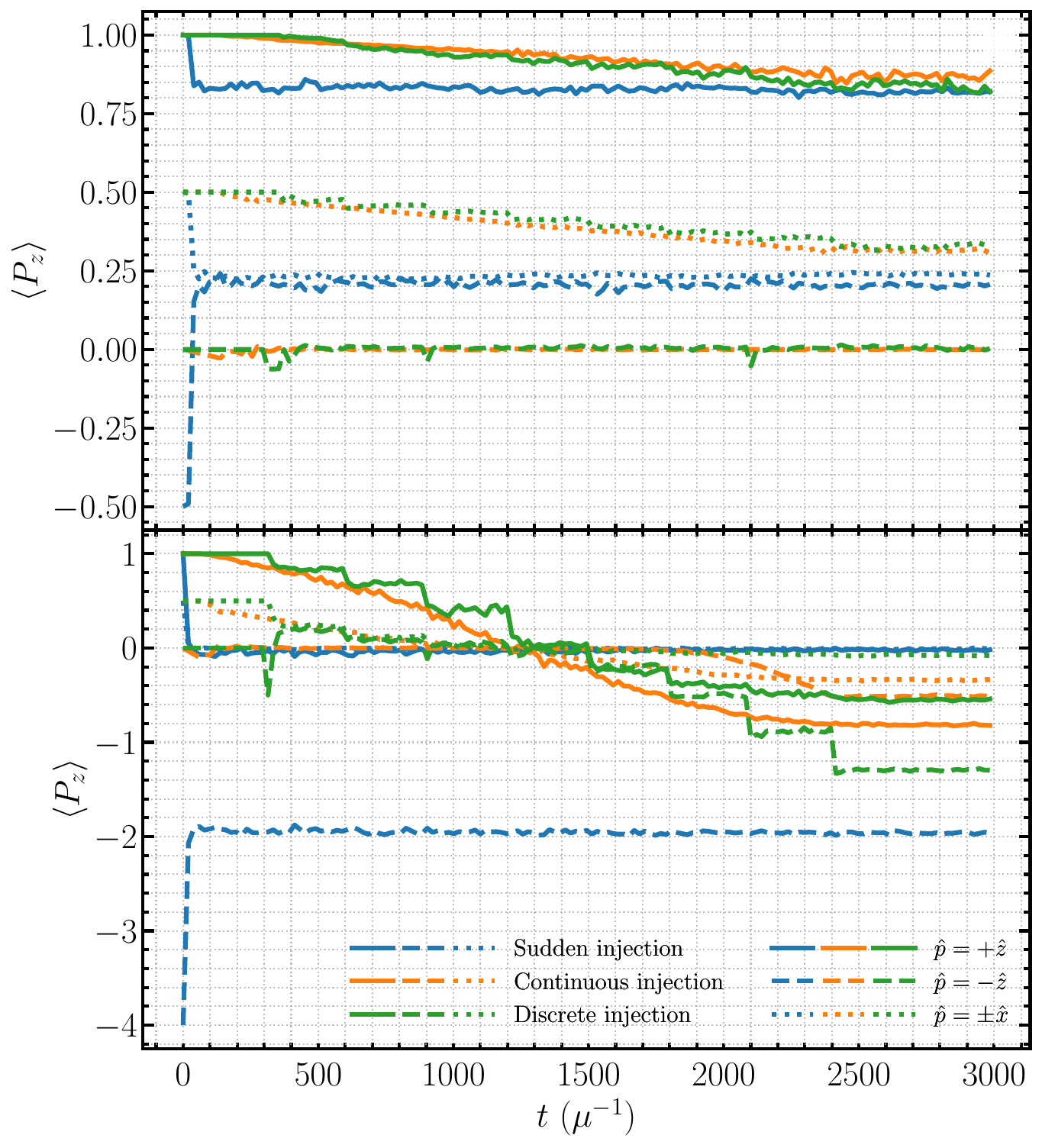}
    \caption{\label{fig:seqins} 
        We test effective classical transport subgrid models by driving fast flavor instabilities through three different mechanisms: sudden, discrete, and continuous injection. In the sudden case, the system is initialized directly in an unstable configuration. In the discrete case, the instability is triggered by injecting eight heavy-neutrino packets, while in the continuous case, it develops through sustained heavy-neutrino injection. In all cases, the injection occurs exclusively in the dashed beam propagating along the $-z$ direction. Simulations in the lower panel experience eight times more injection than those in the upper panel. The distinct driving schemes lead to different final flavor compositions, confirming that the way the instability is driven influences the outcome of flavor evolution. The observed discrepancies highlight the limitations of the two-step approach, in which instabilities arise abruptly and subsequently relax. Such simplified procedures fail to fully capture the nonlinear dynamics of flavor conversion. The simulation legends apply to both upper and lower panels.
    }
    \end{figure}

    In all simulations of Fig.~\ref{fig:seqins}, the beams going in the $+z$ and $\pm x$ directions have number densities as described in Sec.~\ref{sec:numsetup}. The difference is in the injection scheme in the beam going in the $-z$ direction, which can be sudden, discrete, or continuous. The sudden-injection cases (blue) are initialized directly in the unstable state with number densities of heavy neutrinos of $n^{+z}_{\mathrm{init}}/2$ for the upper panel and $4n^{+z}_{\mathrm{init}}$ for the lower panel (eight times more than the former) propagating in the $-z$ direction, respectively. In the discrete-injection case, the beam propagating in the $-z$ direction is initialized with zero electron and heavy flavor neutrinos. The instability is triggered by the injection of eight heavy-neutrino packets, each with a number density of $n^{+z}_{\mathrm{init}}/(2\times8)$ (upper panel) and $4n^{+z}_{\mathrm{init}}/8$ (lower panel). The packets are injected every $(2400/8)\,\mu^{-1}$ within the interval $0 \le t \le 2400\,\mu^{-1}$. For the continuous injection case the beam propagating in the $-z$ direction is also initialized with zero electron and heavy flavor neutrinos. The injection term for this beam is given by
    \begin{eqnarray}
        \dot{n}_{ab}^{-z} = \frac{n^{+z}_{\mathrm{init}}}{2}\frac{\delta_{ax}\delta_{bx}}{2400\,\mu^{-1}}
        \hspace{2mm},\hspace{2mm} 0 \le t \le 2400\,\mu^{-1}
        \label{eq:cont_col_term_1xemission}
    \end{eqnarray}
    for the upper panel, and
    \begin{eqnarray}
        \dot{n}_{ab}^{-z} = 4n^{+z}_{\mathrm{init}}\frac{\delta_{ax}\delta_{bx}}{2400\,\mu^{-1}}
        \hspace{2mm},\hspace{2mm} 0 \le t \le 2400\,\mu^{-1}
        \label{eq:cont_col_term_8xemission}
    \end{eqnarray}
    for the lower panel. $a$, $b$ $\in \{e,x\}$ where $e$ is electron flavor and $x$ represent heavy flavor. This implies that electron neutrino injection is set to zero in this beam. Beams in other directions do not experience injection. Antineutrinos are not included.
    
    The standard FFC prescription in Eqs.~\ref{eq:Pij1} and \ref{eq:Pij2} requires the system to eliminate the ELN-XLN crossing on its shallow side, while other angular regions adjust their flavor conversion to maintain conservation laws \cite{zaizen2023simple}. In the sudden case in the upper panel of Fig.~\ref{fig:seqins}, the shallow-side crossing (blue dashed beam initially in negative $P_z$) is not eliminated but instead overshoots into positive values. In contrast, the sudden case in the lower panel of Fig.~\ref{fig:seqins} successfully removes the ELN-XLN crossing in the shallow side (blue solid and dotted beams initially at positive $P_z$).
    
    The generation of ELN-XLN crossing via discrete (green) or continuous (orange) injection drives the system toward a final state distinct from the sudden cases. This demonstrates error due to the two-step nature of effective classical transport. This also supports the arguments presented in Ref.~\cite{johns2024subgrid} and, in particular, the numerical findings of Ref.~\cite{fiorillo2024fast}. The latter study showed explicitly that the manner in which the instability is induced---through sudden versus continuous driving---affects the final flavor content in some cases. These results affirm the necessity of a self-consistent, continuous flavor-mixing response in neutrino-radiation environments, such as core-collapse supernovae and neutron star mergers, in order to correctly capture the associated fluid dynamics and subsequent astrophysical observables impacted by neutrinos.
    
    The influence of the instability driving scheme becomes more pronounced in the lower panel of Fig.~\ref{fig:seqins}, where the heavy-neutrino injection rate is increased eightfold. In the discrete case (green), once the first packet is injected into the dashed beam ($t = 300\,\mu^{-1}$), $P_z$ quickly overshoots zero and becomes positive, again showing that the instability does not necessarily saturate by eliminating the ELN-XLN angular crossing. In contrast, the continuous injection case (orange) quasi-statically drives the crossing toward zero. All beams reach $\langle P_{z} \rangle = 0$ at $\approx 1300\,\mu^{-1}$. Beyond this point, heavy-neutrino injection no longer induces an ELN-XLN crossing, yet flavor conversion continues in both the discrete and continuous scenarios (we will discuss this phenomenon in Sec.~\ref{sec:homog}). Beams propagating along the $\pm x$ and $+z$ directions not only eliminate the crossing but overshoot into negative values, making their final states distinct from the sudden case, which merely removes the crossing. Moreover, the asymptotic states resulting from the discrete and continuous injection also differ from each other, further demonstrating that the way the instability is driven influences the final flavor composition. This discrepancy is associated with the two-step nature of the process, in which instabilities are suddenly generated and then relaxed. This raises concerns about the reliability of the two-step procedure of effective classical transport. 

\section{Subgrid homogenization}
    \label{sec:homog}
    
    In this section, we examine how flavor inhomogeneities affect subsequent flavor evolution when instabilities are generated continuously and the neutrino flavor field is evolve quasi-statically. Our goal is to test whether subgrid flavor inhomogeneities play a significant role in astrophysical environments such as core-collapse supernovae and binary neutron star mergers. In these scenarios, instabilities that generate flavor inhomogeneities may arise continuously. Nevertheless, all asymptotic-state prescriptions proposed so far remain spatially homogeneous at subgrid scales (e.g., \cite{zaizen2023simple, xiong2023evaluating}). We show that neglecting inhomogeneities in subgrid treatments leads to discrepancies.

\begin{figure}[!htbp]
        \centering
        \includegraphics[width=0.983\linewidth]{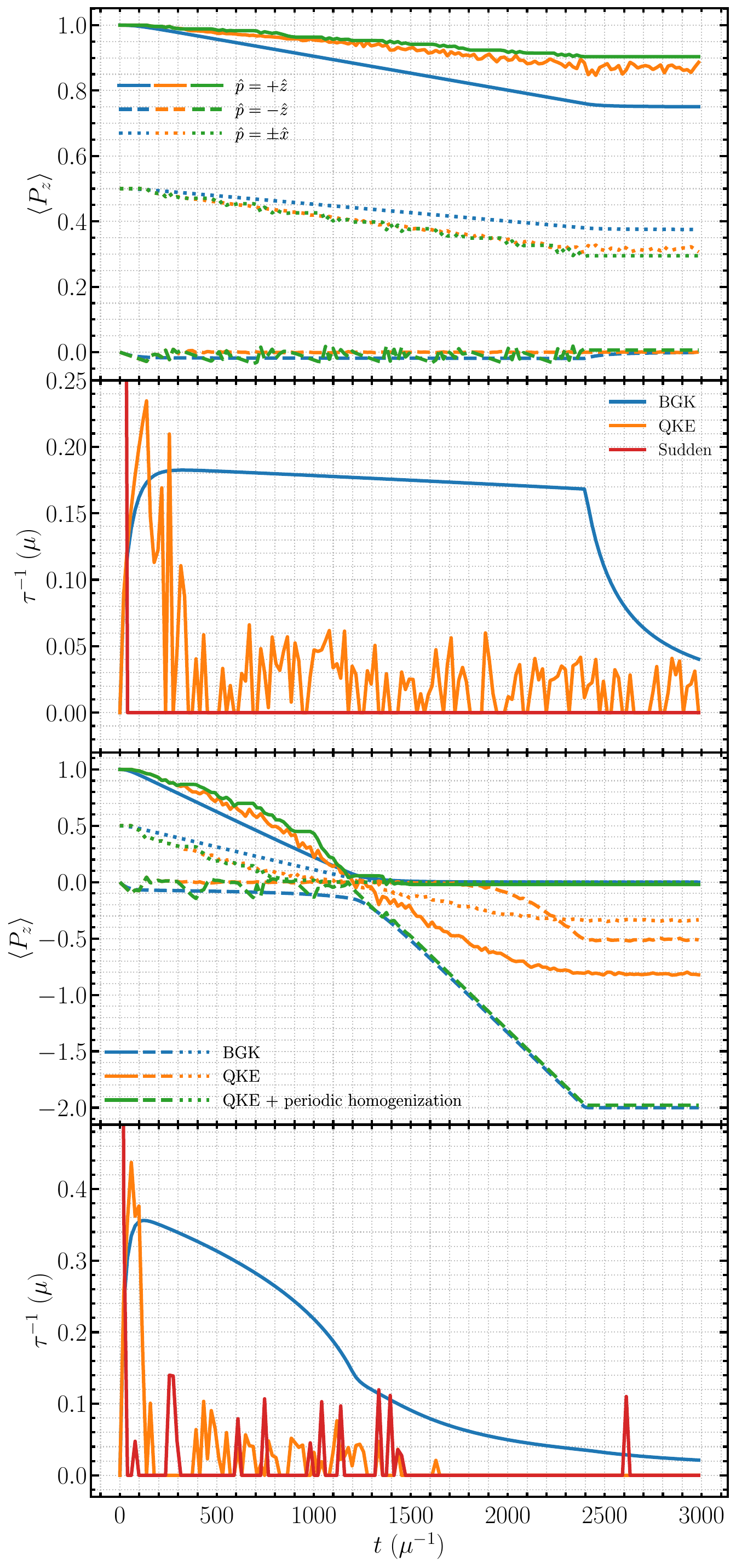}
        \caption{\label{fig:subgrid_homogenization} 
            Comparison of homogenous flavor conversion paths (BGK and QKE + periodic homogenization) versus inhomogeneous (QKE). The relaxation time $\tau(t)$ is shown for BGK, QKE, and sudden injection (blues in Fig.~\ref{fig:seqins}). Simulations in the lower panels experience eight times more injection than those in the upper panel. See Sec.~\ref{sec:homog} for more simulation details. Imposing subgrid homogeneity alters the dynamics, producing large deviations from the behavior predicted by the QKE evolution. FFC can occur without instability or ELN-XLN angular crossing (see QKE bottom panels for $t>1200\,\mu^{-1}$). The emergent relaxation timescale $\tau$ for flavor conversion is larger when the system is driven quasi-statically compare to the sudden cases where $\tau\sim\mu^{-1}$. The simulation legends apply to all panels.
            }
    \end{figure}

    We perform three different simulations: QKE, QKE + periodic homogenization, and BGK. They are shown in Fig.~\ref{fig:subgrid_homogenization}. In all cases, the beams propagating in the $+z$ and $\pm x$ directions have number densities as described in Sec.~\ref{sec:numsetup}. The beam propagating in the $-z$ direction is initialized with zero electron and heavy-flavor neutrinos. The heavy-neutrino injection follows Eq.~\ref{eq:cont_col_term_1xemission} (upper panels) or Eq.~\ref{eq:cont_col_term_8xemission} (lower panels).
    
    The QKE (orange) represents the inhomogeneous quantum-kinetic solution (identical to the continuous-injection case in Fig.~\ref{fig:seqins}). The QKE + periodic homogenization (green) resemble the QKE evolution but with an imposed homogenization applied eight times, every $(2400/8)\,\mu^{-1}$ within the interval $0 \le t \le 2400\,\mu^{-1}$ so flavor inhomogeneities develop only briefly before the system is re-homogenized. At each homogenization step, the polarization vector $\mathbf{P}$ is reset to its domain-averaged value, after which a random perturbation of amplitude $10^{-4} P_z$ is introduced in $P_\perp$. We also show the BGK solution (blue), which is described in Sec.~\ref{sec:bgk}.
    
    Our intention with the QKE + periodic homogenization and BGK simulations is to mimic an effective classical transport implementation where  
    (1) the neutrino field is driven toward instability;  
    (2) the resulting unstable neutrino distribution is passed to a flavor conversion solver, which allows inhomogeneous flavor waves to develop (QKE + periodic homogenization allows inhomogeneous but not BGK);  
    (3) The output of the flavor-conversion solver is then fed back into step (1), but without the spatial flavor structures generated by the instabilities (this corresponds to the times when periodic homogenization is applied in the QKE + periodic homogenization simulation).
    This process is then repeated. 
    In the BGK model, the neutrino flavor distributions are continuously driven toward an asymptotic state like in (3), with subgrid inhomogeneity erased.
    By comparing the QKE, QKE with periodic homogenization, and BGK solutions, we can quantify the error introduced when subgrid models neglect flavor inhomogeneities.
    
    The final flavor content of the inhomogeneous QKE differs from BGK and QKE + periodic homogenization. This is especially evident in the lower panels of Fig.~\ref{fig:subgrid_homogenization}. In this panel before the beams cross (up to $1200\,\mu^{-1}$), all simulations quasi-statically respond to the instability, removing the ELN--XLN crossing. After the beams cross, there is no further generation of ELN--XLN crossing; nonetheless, the inhomogeneous QKE continues to undergo flavor conversion. This flavor conversion cannot be explained by FFI or the standard ELN--XLN crossing criterion and provides evidence that flavor conversion can occur without an instability. Moreover, since both BGK and QKE with periodic homogenization, where the spatial structures were erased, show no further conversion, the observed convertion in QKE must be intrinsically tied to the spatial flavor structures developed by previous instabilities. Thus, the imposition of subgrid homogeneity engenders great discrepancies with the QKE evolution.
    
    The BGK simulation fails to reproduce the QKE flavor trajectory when fast inhomogeneous instabilities are continuously driven. This is more noticeable in the bottom two panels of Fig.~\ref{fig:subgrid_homogenization}. In this sense, BGK and QKE + periodic homogenization are equivalent: both eliminate the ELN--XLN crossing and both fail to capture the flavor conversion observed in the inhomogeneous QKE simulation, particularly where FFC occurs in the absence of ELN--XLN crossings. Although the BGK subgrid model resolves the continuous/sequential problem noted in Sec.~\ref{sec:sequential}, it imposes a local relaxation toward a homogeneous asymptotic state, leading to deviations from the full QKE evolution.
    
    Fig.~\ref{fig:subgrid_homogenization} also shows the relaxation timescale $\tau$ defined in Eq.~\ref{eq:tau} for the BGK, QKE, and Sudden (blue in Fig.~\ref{fig:seqins}) simulations. This timescale emerges naturally from the competition between the instability-driving mechanism and the flavor conversion that relaxes this instability. As expected, the sudden cases have initial relaxation timescales $\tau\sim\mu^{-1}$, being $0.99\,\mu^{-1}$ (up) and $2.79\,\mu^{-1}$ (low), respectively. However, when the system is driven quasi-statically, with the instability driven by continuous injection, the emergent timescale is smaller (see QKE and BGK).  BGK initially agrees with the QKE relaxation timescale but quickly diverges and underestimates it. However, this inconsistency does not necessarily rule out the BGK framework itself. It may instead reflect the limited accuracy with which we estimate the relaxation time $\tau$ and the asymptotic distribution in the approximate scheme of Sec.~\ref{sec:bgk}. With a more accurate prescription for these quantities, the BGK model could, in principle, reproduce the QKE results. Multiple efforts have been made to develop reliable asymptotic state models, including approaches based on machine learning \cite{richers2024asymptotic} and other analytical prescriptions \cite{liu2025asymptotic, gdg9-rzns}, but further investigation is required to shed light on realistic flavor relaxation timescales and asymptotic states considering the combined effects of the instability-driving mechanisms and the quasi-static flavor conversion processes that counteract them.

\section{Reversibility\label{sec:irrev}}
    
    We now investigate the reversibility of a kinetic neutrino gas undergoing FFC by running four simulations that in the absence of flavor conversion return to the initial states. These simulations show especially clearly that FFC can occur without being accompanied by flavor instability. This is the third limitation we highlight: Effective classical transport and the BGK model are limited by the fact that they have exclusively been implemented using asymptotic states resulting from driving into unstable parameter space. We show explicitly that FFC can occur due to driving through stable parameter space as well.
    
\begin{figure}
        \includegraphics[width=1.0\linewidth]{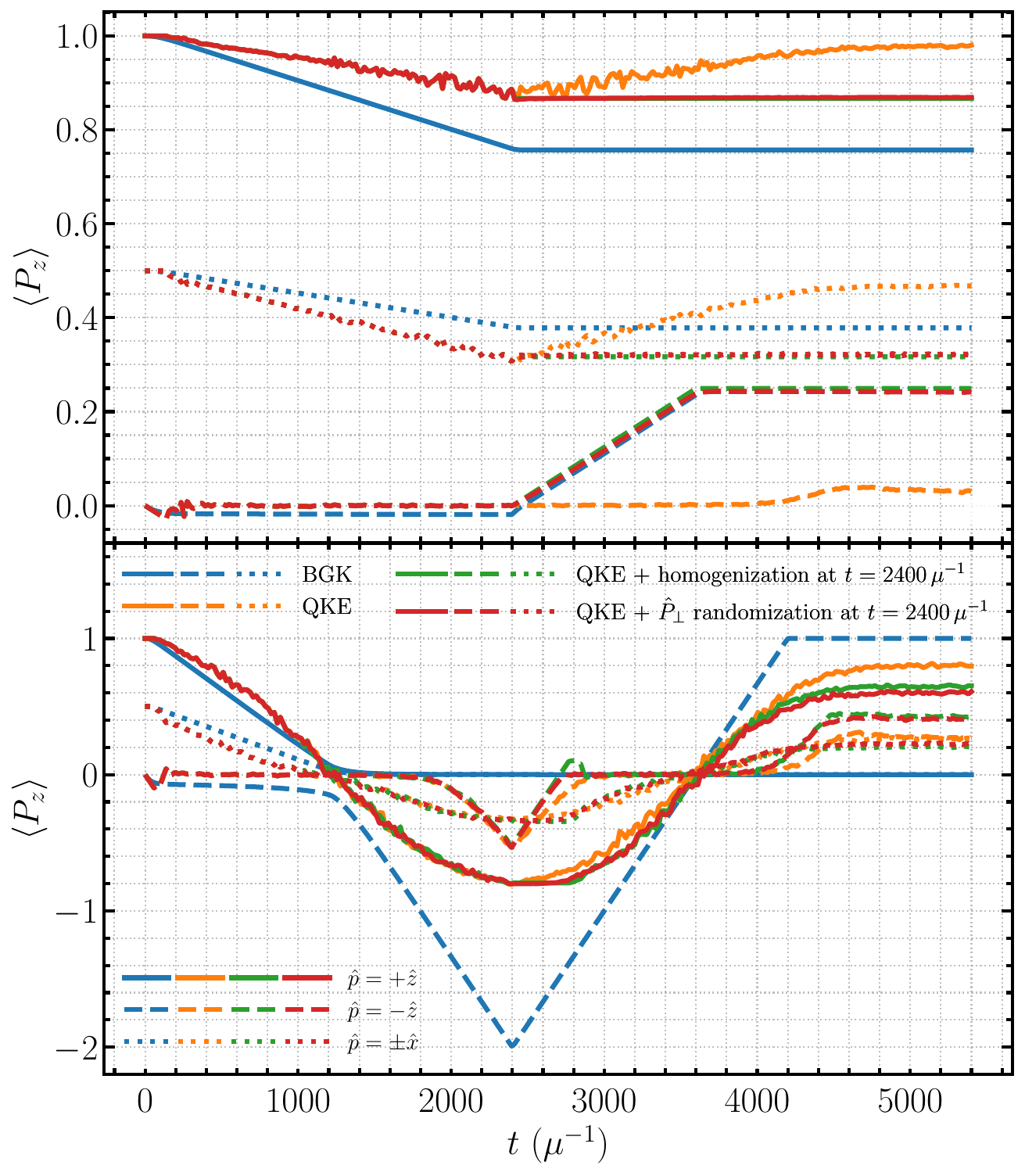}
        \caption{\label{fig:reversibility}
            Reversibility properties of a kinetic neutrino gas undergoing FFC. ELN-XLN crossing is induced by continuous injection of heavy neutrinos in the dashed beam followed by removal the same amount (or until there is not more heavy neutrinos). Simulations in the lower panel experience eight times more injection than those in the upper panel. See injection and removal term in Eq.~\ref{eq:cont_col_term_1xemission_rever} for upper panel and Eq.~\ref{eq:cont_col_term_8xemission_rever} for lower panel.
            In the absence of flavor conversion, the system is reversible since it returns to its initial state. 
            QKE + homogenization at $t = 2400\,\mu^{-1}$ reset $\mathbf{P}$ to its domain-averaged seeding a random perturbation of amplitude $10^{-4} P_z$ in $P_\bot$.
            QKE + randomization at $t = 2400\,\mu^{-1}$ randomly rotates $P_\bot$ around $P_z$.
            The upper ($t\in[2400,4800]\,\mu^{-1}$) and lower ($t\in[1200,2400]\,\mu^{-1}$ and $t\in[3600,4800]\,\mu^{-1}$) panels show that FFC can occur without being accompanied by flavor instability and ELN-XLN crossing. The simulation legends apply to both upper and lower panels.
        }
    \end{figure}

    We perform four different simulations: QKE, QKE + homogenization at $t=2400\,\mu^{-1}$, QKE + randomization at $t=2400\,\mu^{-1}$ and BGK. They are shown in Fig.~\ref{fig:reversibility}. In all cases, the beams propagating in the $+z$ and $\pm x$ directions have number densities as described in Sec.~\ref{sec:numsetup}. The beam propagating in the $-z$ direction is initialized with zero electron and heavy-flavor neutrino content, and the instability, namely the ELN--XLN crossing, is induced through the continuous injection of heavy neutrinos into that beam.
    This is followed by the removal of an equal amount or until there is not more heavy neutrinos, as described by the following equation
    \begin{eqnarray}
    \dot{n}_{ab}^{-z}(t)=
    \begin{cases}
    +\dfrac{n^{+z}_{\mathrm{init}}}{2}\,\dfrac{\delta_{ax}\delta_{bx}}{2400\,\mu^{-1}}
    \hspace{2mm},\hspace{2mm} 0 \le t \le 2400\,\mu^{-1}\\[12pt]
    -\dfrac{n^{+z}_{\mathrm{init}}}{2}\,\dfrac{\delta_{ax}\delta_{bx} }{2400\,\mu^{-1}}
    \hspace{2mm},\hspace{2mm} 2400 < t \le 4800\,\mu^{-1}
    \end{cases}
    \label{eq:cont_col_term_1xemission_rever}
    \end{eqnarray}
    for the top panel, and
    \begin{eqnarray}
    \dot{n}_{ab}^{-z}(t)=
    \begin{cases}
    +4\,n^{+z}_{\mathrm{init}}\,\dfrac{\delta_{ax}\delta_{bx}}{2400\,\mu^{-1}}
    \hspace{2mm},\hspace{2mm} 0 \le t \le 2400\,\mu^{-1}\\[12pt]
    -4\,n^{+z}_{\mathrm{init}}\,\dfrac{\delta_{ax}\delta_{bx}}{2400\,\mu^{-1}}
    \hspace{2mm},\hspace{2mm} 2400 < t \le 4800\,\mu^{-1}
    \end{cases}
    \label{eq:cont_col_term_8xemission_rever}
    \end{eqnarray}
    for the lower panel. $a$, $b$ $\in \{e,x\}$ where $e$ is electron flavor and $x$ represent heavy flavor. During the injection and removal phases, the flavor evolution remains quasi-static. In the absence of flavor conversion, the system is reversible, meaning it returns to its initial state. The QKE curve (orange) represents the inhomogeneous quantum-kinetic solution. At the transition between injection and removal ($t = 2400\,\mu^{-1}$), we study the flavor evolution when the spatial flavor structure is homogenized (green: QKE + homogenization at $t = 2400\,\mu^{-1}$). In this homogenization, the polarization vector $\mathbf{P}$ is reset to its domain-averaged value, after which a random perturbation with amplitude $10^{-4} P_z$ is introduced in $P_\perp$. We also examine the case where the transverse polarization vectors $P_\perp$ are randomized (red: QKE + randomization at $t = 2400\,\mu^{-1}$). In this simulation, $P_\perp$ is randomly rotated around $P_z$ while preserving the vector magnitude. The BGK evolution (blue) is shown for comparison.
    
    In the upper panel of Fig.~\ref{fig:reversibility} for $t\in[0,2400]$, QKE and QKE + homogenization quasi-statically zeroes out the ELN--XLN crossing generated by the heavy-neutrino injection in the dashed beam. In this process, not only is the angular distribution affected, but spatial flavor structure is also generated. The BGK solution also zeroes out the crossing but fails to generate this spatial flavor structure and converges to a different asymptotic state. At $t=2400\,\mu^{-1}$, the dynamics switches from injection to removal of heavy neutrinos in the dashed beams. Although removal does not induce an ELN--XLN crossing nor instabilities, QKE still exhibits flavor conversion. The removed heavy neutrinos in the dashed beam are balanced by transferring electron neutrinos to other angular directions. This flavor conversion can be rooted in the spatial flavor structure generated by previous instabilities. When the system is homogenized in QKE + homogenization and BGK or QKE + $P_\perp$ randomization, the condition that leads to flavor conversion disappears, and the system is driven solely by the heavy-neutrino removal. Homogeneous subgrid models fail to capture flavor conversion arising from the spatial flavor structure legacy of previous instabilities.
    
    The same number of heavy neutrinos are injected and removed, although in simulations with flavor conversion they may be completely removed before reaching the number originally injection. The classical solution is reversible in the sense that the final and initial states are identical. There is a tendency in the QKE solution to maintain reversibility, but this is not completely achieved. QKE + homogenization, BGK and QKE + $P_\bot$ randomized are not reversible at all.
    
    In the lower panel for $t\in[1200,2400]$, the QKE and QKE + homogenization exhibit behavior similar to the QKE in the upper panel for $t\in[2400,4800]$, but with $\langle P_z \rangle$ shifted to the negative side. Although the crossings vanish around $t=1200\,\mu^{-1}$, flavor conversion continues, showing that this process does not require ELN-XLN crossings. This demonstrates that the behavior in the upper panel is not triggered by the injection-removal switch, but by a beam crossing in a flavor spatially inhomogeneous environment. Specifically, in the upper panel the dashed (injecting/removing) beams cross, while in the lower panel the crossing occurs between the solid and dotted (non-injecting) beams.
    
    At $t=2400\,\mu^{-1}$ in the lower panel of Fig.~\ref{fig:reversibility}, the dashed beams switch from injection to removal. QKE + homogenization, having lost spatial flavor structure, ceases flavor conversion, whereas QKE continues flavor conversion even when no ELN-XLN crossing exists but flavor structure imprinted earlier. Around $t=2800\,\mu^{-1}$, removal induces an ELN-XLN angular crossing in the QKE + homogenization simulation, allowing flavor conversion to resume and regenerate spatial structure. Between $t\in[2800,3600]$, all simulations except BGK develop crossings and subsequently eliminate them, as expected.
    
    At approximately $t=3600\,\mu^{-1}$, all beams cross to positive $\langle P_z \rangle$ values. Although removal in the dashed beams no longer induces ELN-XLN crossings, QKE + homogenization, BGK, and QKE + $P_\perp$ randomization still exhibit flavor conversion. This conversion originates from the spatial flavor structure imprinted earlier by the crossings at $t=2800\,\mu^{-1}$. The asymptotic states of QKE and QKE + homogenization differ, underscoring the key role of spatial structure in shaping the long-term flavor evolution.
    
    The BGK solution, inherently homogeneous, cannot capture this structure-driven conversion. Its asymptotic state depends solely on ELN-XLN–induced conversion, leading to inaccuracies. Incorporating spatially inhomogeneous asymptotic states could improve BGK’s predictive power, especially under the generation of continuous instabilities where the system evolves quasi-statically.
    
\section{Subgrid phase information\label{sec:random}}
    
    In this section, we study flavor evolution when subgrid phase information is periodically randomized. At this point, it remains unclear which subgrid details are essential for a reliable coarse-grained implementation of FFC and which can be safely discarded. If no information can be neglected, the viability of all coarse-grained approaches becomes questionable. Motivated by this, we investigate whether quantum phases play a significant role in determining the flavor evolution.
    
\begin{figure}
    \includegraphics[width=1.0\linewidth]{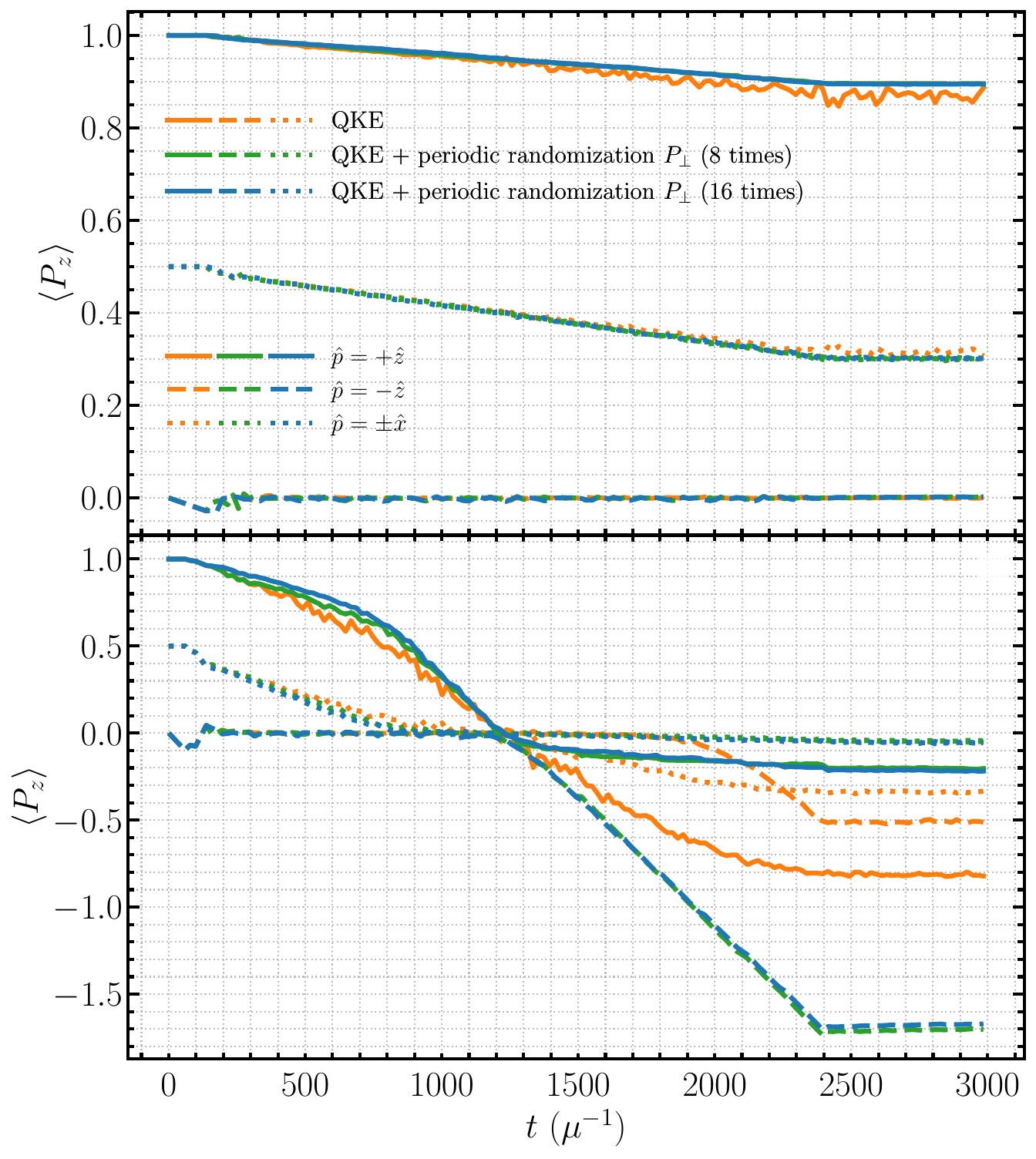}
    \caption{\label{fig:turbulence_viscovity_1xem}
        Flavor evolution when subgrid phase information at each location and direction is randomized by rotating the polarization vector $\mathbf{P}$ by a random phase around the $P_z$ axis, while keeping $|\mathbf{P}|$ and $|P_{\perp}|$ fixed. Simulations in the lower panel experience eight times more injection than those in the upper panel. See Sec.~\ref{sec:random} for more simulation details. The randomization is applied periodically (8 and 16 times). Within the first $2400\,\mu^{-1}$, an ELN--XLN crossing is induced by injecting $n^{+z}_{\mathrm{init}}/2$ (top) and $4n^{+z}_{\mathrm{init}}$ (bottom) heavy neutrinos in the dashed $-z$ beam.  See injection term in Eq.~\ref{eq:cont_col_term_1xemission_rever} for upper panel and Eq.~\ref{eq:cont_col_term_8xemission_rever} for lower panel. When subgrid phase information is lost, flavor conversion without ELN-XLN crossing and FFI still occurs but deviates from the QKE evolution, indicating that quantum-coherent phases provide essential feedback in the subsequent flavor dynamics. The simulation legends apply to both upper and lower panels.
        }
    \end{figure}

    We perform three different simulations: QKE, QKE + periodic randomization $P_{\bot}$ (8 times), and QKE + periodic randomization $P_{\bot}$ (16 times). They are shown in Fig.~\ref{fig:turbulence_viscovity_1xem}. In all cases, the beams propagating in the $+z$ and $\pm x$ directions have number densities as described in Sec.~\ref{sec:numsetup}. The beam propagating in the $-z$ direction is initialized with zero electron and heavy-flavor neutrinos and receives a continuous heavy-neutrino injection following Eq.~\ref{eq:cont_col_term_1xemission} (upper panel) and Eq.~\ref{eq:cont_col_term_8xemission} (lower panel).
    
    The QKE (orange) represents the inhomogeneous quantum-kinetic solution (same as the continuous-injection case in Fig.~\ref{fig:seqins} and QKE in Fig.~\ref{fig:subgrid_homogenization}). 
    The QKE + periodic randomization $P_{\bot}$ (8 times, green) corresponds to the same inhomogeneous quantum-kinetic evolution but with an imposed rotation of the polarization vector $\mathbf{P}$ by a random phase around the $P_z$ axis. 
    In this process, we keep $|\mathbf{P}|$ and $|P_\bot|$ fixed, while only the direction of $P_{\bot}$ is randomly reassigned. 
    The resulting state is further evolved. 
    This randomization is applied eight times, every $(2400/8)\,\mu^{-1}$ within the interval $0 \le t \le 2400\,\mu^{-1}$. The QKE + periodic randomization $P_{\bot}$ (16 times) is identical to the former but the randomization is applied sixteen times, every $(2400/16)\,\mu^{-1}$ within the interval $0 \le t \le 2400\,\mu^{-1}$. 
    We also show the BGK solution (blue), which is described in Sec.~\ref{sec:bgk}.
    
    In the upper panel of Fig.~\ref{fig:turbulence_viscovity_1xem}, the instability induces flavor conversion across all beams, working to erase the shallow ELN--XLN crossing. The resulting asymptotic flavor states differ between the QKE and periodic randomization in $P_{\perp}$, indicating that subgrid phase information plays a key role in determining the flavor evolution path. In the lower panel, where the system experiences eight times higher heavy-neutrino injection, the beams cross $\langle P_{z} \rangle = 0$ at $t \approx 1400\,\mu^{-1}$. Beyond this point, no ELN--XLN crossing or FFI remains, and the subsequent flavor conversion originates from the spatial flavor structure generated by previous instabilities. Randomizing the polarization vector components perpendicular to the $P_z$ axis reduces the amount of flavor conversion (green and blue compared to orange). As shown previously, complete homogenization of the system removes any flavor conversion. When subgrid phase information is erased, flavor conversion still occurs but follows a distinct path compared to the QKE evolution, highlighting the crucial role of the spatial structure of quantum coherence in the subsequent flavor dynamics.
    
    Some directions that transition between positive and negative $\langle P_z \rangle$, such as at $t \approx 1400\,\mu^{-1}$ in the lower panel of Fig.~\ref{fig:turbulence_viscovity_1xem}, amplify the impact of spatial flavor inhomogeneities and subgrid phase information on the subsequent flavor evolution. Beam crossings appear to play a critical role in continuously driven FFC \cite{fiorillo2024fast, liu2024quasi}. A detailed investigation of this condition will be pursued in future work.

\section{Summary\label{sec:summary}}
    
    Neutrino subgrid models of two types have so far been implemented into astrophysical simulations: effective classical transport \cite{li2021, xiong2024robust} and the neutrino BGK model \cite{nagakura2024bhatnagar}. These coarse-grained approaches have the virtue of being computationally tractable. However, they are of course simplifications of the full flavor-mixing physics, and there is room for further improvement. In this study we have highlighted certain key limitations of effective classical transport and the BGK model. We hope our calculations and discussions will inspire future work aimed at addressing these shortcomings.
    
    The first limitation, which is faced by effective classical transport but not the BGK model, is that the approximation of simultaneous driving and flavor response by a two-step sequence (driving followed by response) introduces errors. These errors can be mitigated by adopting a smaller step size. They can also be avoided by using the BGK model, wherein neutrino flavor continuously relaxes to a local, instantaneous asymptotic state.
    
    The second limitation is the restriction, up to this point, of effective classical transport and the BGK model to spatially homogeneous asymptotic states. Flavor instabilities cause inhomogeneous perturbations to grow. Spatially homogeneous asymptotic states discard this subgrid information, but as we have shown, small-scale inhomogeneity created during prior evolution may significantly affect subsequent flavor conversion. It is important to note that the periodic bondary conditions used tend to enhance and preserve flavor inhomogeneities. In reality, flavor inhomogeneities can be depleted by natural mechanisms such as advection and collisions that damp the small-scale structure. The role of these processes in continuously triggering instabilities, as in our calculations, is still an open question, and further effort is needed to address this.
    
    The influence of subgrid inhomogeneity is particularly notable when neutrinos are driven through stable states. We have demonstrated that FFC can occur even in the absence of instabilities. Thus the third limitation is that effective classical transport and the BGK model have so far only been implemented using post-instability asymptotic states. This issue is intertwined with the previous one because stable FFC hinges on inhomogeneity.
    
    The subgrid methods we have focused on could conceivably be extended to permit subgrid inhomogeneity. The mapping onto asymptotic states could perhaps also be broadened to encompass asymptotic states resulting from any type of driving, not only driving into unstable regions of parameter space. The concern here is that the mapping potentially becomes far more complex once inhomogeneity is allowed in initial and asymptotic states. These ideas warrant further investigation.
    
\begin{acknowledgments}

    The authors gratefully acknowledges Sherwood Richers, Irene Tamborra, Manuel Goimil-García, Meng-Ru Wu, Hiroki Nagakura, Damiano Fiorillo, Georg Raffelt and Hans-Thomas Janka for their valuable feedback on this work. L.J. is supported by a Feynman Fellowship through LANL LDRD project number 20230788PRD1. E.U. is supported by the Dr. Elizabeth M. Bains and Dr. James A. Bains Graduate Research Fellowship of the Physics and Astronomy Department of the University of Tennessee, Knoxville.

\end{acknowledgments}

\bibliography{refs}
\bibliographystyle{apsrev4-2}

\end{document}